# The influence of Jupiter, Mars and Venus on Earth's orbital evolution

Jonathan Horner[1,2], James B. Gilmore[2], and Dave Waltham[3]

[1] Computational Engineering and Science Research Centre, University of Southern Queensland, Toowoomba, Queensland 4350, Australia
[2] Australian Centre for Astrobiology, UNSW Australia, Sydney, New South Wales 2052, Australia
[3] Department of Earth Sciences, Royal Holloway, University of London

**Summary:** In the coming years, it is likely that the first potentially Earth-like planets will be discovered orbiting other stars. Once found, the characterisation of those planets will play a vital role in determining which will be chosen as the first targets for the search for life beyond the Solar System. We must thus be able to gauge the relative importance of the various factors proposed to influence potential planetary habitability, in order to best focus that search.

One of the plethora of factors to be considered in that process is the climatic variability of the exo-Earths in question. In the Solar System, the Earth's long-term climate is driven by several factors, including the modifying influence of life on our atmosphere, and the temporal evolution of solar luminosity. The gravitational influence of the other planets in the Solar System adds an extra complication, driving the Milankovitch cycles that are thought to have caused the on-going series of glacial and interglacial periods that have dominated Earth's climate for the past few million years.

Here we present preliminary results of three suites of integrations that together examine the influence of Solar System architecture on the Earth's Milankovitch cycles. We consider separately the influence of the planets Jupiter, Mars and Venus, each of which contributes to the forcing of Earth's orbital evolution. Our results illustrate how small changes to the architecture of a given planetary system can result in marked changes in the potential habitability of the planets therein, and are an important first step in developing a means by which the nature of climate variability on planets beyond our Solar System can be characterised.

# Introduction

The question of whether we are alone in the universe is one that has long fascinated humankind. In the early to mid-twentieth century, speculation abounded as to the possibility of complex and advanced life on our nearest planetary neighbours – Mars and Venus (e.g. [1][2][3]). Some authors even went so far as to speculate on the forms that Martian and Venusian life might take (e.g. [4]).

With the dawn of the space age, and the advent of the first missions to other planets, it soon became abundantly clear that neither Mars nor Venus could host advanced life[1]. The Mariner 9 spacecraft revealed Mars to be an arid world, cold and desolate (e.g. [5][6]), whilst the Venera missions showed Venus to be less habitable still, with a surface hot enough to melt lead (e.g. [7][8]). The observations of Mars suggested a past that might once have been wetter and warmer than the present (e.g. [9]), but neither planet offered hope for the discovery of contemporaneous advanced life.

In the past two decades, the question of life elsewhere has once again become the topic of serious discussion. Within our Solar System, the discovery of liquid water in a variety of locations has reawakened the possibility that microscopic life could find habitats beyond the Earth. Observations taken by the Galileo orbiter indicate the presence of sub-surface oceans on the Jovian satellites Europa ([10][11]), Ganymede ([12]), and Calisto ([13][14]). Similarly, and more surprisingly, the Cassini orbiter has found striking evidence of liquid water on the small Saturnian moon, Enceladus (e.g. [15][16]). Once again, however, the most promising target for the search for life in the Solar system seems to be Mars, with recent observations confirming the presence of water-based salty liquid brine on the planet's surface ([17]). However, despite this, there remains some debate over the modern habitability of Mars, as a consequence of its harsh surface conditions (e.g. [91]).

At the same time, we have seen the discovery of the first exoplanets. At first, in the mid-to-late 1990s, the planets found were nothing like the Earth – behemoths comparable in mass to, or more massive than, Jupiter, but orbiting at a fraction of Jupiter's orbital radius (e.g. [18][19][20]). As time has passed, however, the techniques used to search for planets have improved, and the temporal baselines of the searches have grown longer. Taken in concert, these advances have allowed astronomers to discover more distant giant planets, analogous to those in our own Solar System (e.g. [21][22][23]), and also to find ever smaller planets orbiting stars like the Sun (e.g. [24][25][26]).

As a result of the new techniques and instruments available to astronomers, including the exceptionally productive Kepler spacecraft, it is becoming increasingly clear that small planets far outnumber larger ones (e.g. [27][28][29]. A number of new exoplanet search programs will soon begin that will build on this knowledge, greatly increasing the number of small exoplanets known. In the next decade, new space-based surveys (the Kepler K2 mission [30][31], TESS [32], and PLATO [33]) will be complemented by observations taken by dedicated ground based facilities, such as MINERVA ([34]), NRES ([35]), and the newly announced MINERVA-Australis, to be constructed in 2016 at the University of Southern Queensland's Mt. Kent Observatory.

In the coming years, it is likely that these new exoplanet search programs will discover the first truly Earth-like planets orbiting nearby stars – planets of comparable size and mass to our own, orbiting at a distance from their host star that would allow liquid water to be present and stable on their surface. Once we cross this threshold, the search for life beyond the Solar System will move from the 'needle-in-a-haystack' search for evidence of intelligent aliens (SETI, e.g. [36][37][38]) to the systematic search for biomarkers on known Earth-like planets (e.g. [39][40]). However, the observations needed to characterise the newly discovered planets, and to search for any evidence of life upon them, will be incredibly challenging, and will require highly detailed and time consuming observations (e.g. [41][42]). It will therefore be critically important to determine the most promising targets for such observations, in order to maximise our chances of a positive and timely outcome. But how will that selection process be carried out?

---

[1] In broad terms, advanced life is here taken to mean macroscopic, rather than microscopic life.

Clearly, the ease with which a given target can be observed will play an important role – but beyond that, it is likely that the potential habitability of the planets discovered will be assessed prior to their being chosen for detailed further study. It is thought that a wide variety of factors combine to render one planet more (or less) habitable than another (e.g. [43], and references therein). Among many other factors, the stability of the climate on the planets in question will clearly play an important role in determining the most promising target to study in the search for life elsewhere.

It is well established that the Earth's climate has exhibited significant variability throughout its history. At times, our planet has been far warmer than today (such as during the late Permian and early Triassic periods, some 250 million years ago; [44]), whilst at others it was significantly colder (such as the famous 'Snowball Earth' episodes, thought to have occurred on several occasions through Earth's history – e.g. [45][46]).

These climatic variations have been driven by a wide variety of factors. Earth's climate has been strongly affected by the configuration of the continents (e.g. hot, dry conditions when there was a supercontinent [85]; biological evolution (e.g. substantial draw-down of $CO_2$ from the atmosphere following the first appearance of land-plants [86]); volcanism (e.g. the end-Permian mass-extinction [87]); impacts (e.g. the possible effect of an asteroid collision at the end of the Cretaceous [88][89]); and the steady increase in solar luminosity through time [90].

In addition to these long-term changes, and those driven by stochastic, unpredictable events, the Earth also experiences periodic climatic variation on relatively rapid geological timescales. Perhaps the best demonstration of such variability is that of the last few million years, which have been characterised by a series of glacial and interglacial periods (e.g. [47]). The cause of this semi-periodic behaviour are the Milankovitch cycles (e.g. [48][49]) – variability in the Earth's orbital eccentricity, obliquity and the timing of seasons relative to perihelion passage, driven by the gravitational influence of the Solar System's other planets.

Although the scale of Earth's climatic variability in the last few million years seems relatively dramatic, the variations in Earth's orbit that have driven that variability are actually rather small. The Earth's obliquity (the angle between its equatorial and orbital planes) varies by just over two degrees, whilst its eccentricity rarely ranges higher than ~0.06 (as can be seen in Figure 1 of [54]). It is certainly possible to imagine scenarios where an exo-Earth that might otherwise be highly promising as a target for the search for life would be driven to far greater excursions in eccentricity and inclination than is the Earth, without rendering its orbit sufficiently dynamically unstable as to render the system untenable.

The Milankovitch cycles of the planet Mercury, for example, can drive the planet's orbital eccentricity as high as 0.45 ([50][51]). If that were to be replicated for an otherwise habitable exo-Earth, that planet would experience epochs at which its orbital distance would vary by a factor of almost three from pericentre to apocentre. That would, in turn, result in the amount of flux received by a given region of its surface from its host star varying by almost nine times – something that might preclude, or severely hinder, the development of even the most basic forms of life[2].

Unlike the other processes that drive climate change, therefore, the Milankovitch cycles of a given planet will be unique to that world, driven by the gravitational influence of the other planets in the system. As a result, this offers the potential that the study of the Milankovitch cycles in newly discovered exoplanetary systems could form an important part of a procedure for identifying the most promisingly habitable worlds for further study.

This idea was examined by [52] in the context of our Solar System using a simple analytic approach, examining (among other things) the degree to which the architecture of our Solar System is unusual in the small scale of Earth's Milankovitch cycles. That work suggested that the Earth experiences unusually low Milankovitch frequencies when compared to similar systems with alternative architectures. We note, however, that the analytic model used in that work was unable to account for the influence of resonant effects and direct encounters between planets, and so serves more as the seed of an interesting idea, rather than strong indication of our Solar System's uniqueness. To build on that work, we therefore decided to carry out a similar study that fully models the interactions between the planets involved, as a first step in building a technique to help direct the search for life.

In this paper, we build upon our presentations at the two previous Australian Space Research Conferences ([53][54]), in which we examined the role played by Jupiter in determining the scale and frequency of Earth's Milankovitch cycles. Here, we compare the results of detailed $n$-body integrations that model Jupiter's influence on the amplitude and frequency of Earth's orbital oscillations to new simulations that model the influence of the Earth's nearest planetary neighbours, Mars and Venus.

In the next section, we detail our simulations, recapping those performed to determine Jupiter's influence, and summarising our new runs modelling the effects of Venus and Mars. We then present our preliminary results, before concluding with a discussion of the results and our plans for future work.

## The Simulations

In order to examine how the Earth's Milankovitch cycles change as a function of the Solar System's architecture, we use a modified version of the Hybrid integrator within the $n$-body dynamics package MERCURY ([55]). In its default form, MERCURY neglects relativistic effects, calculating the orbital evolution of objects in a purely Newtonian manner. For most applications (e.g. [56][57][58]), this is no impediment – and performing simulations using a Newtonian rather than relativistic formulism allows the code to run significantly faster than would otherwise be the case. However, the assumption that relativistic effects do not significantly affect the evolution of objects begins to break down as one approaches sufficiently close to massive bodies such as the Sun – as is well illustrated by the failure of Newton's gravitation to explain the precession rate of Mercury's perihelion (e.g. [59][60]).

---
[2] We note that the impact of the Milankovitch cycles would likely be diminished for life in the deep ocean relative to life on land.

Here, we use a modified version of MERCURY that implements a user-defined force to take account of the first-order post-Newtonian relativistic correction ([61]). We have performed tests of the modified code, and have confirmed that it accurately replicates the evolution of Mercury's orbit. The addition of the user-defined force causes the code to run somewhat slower than would otherwise be the case – but thanks to recent advances in the computational power available for us, this has not proved to be too great an impediment.

We have used our code to carry out three suites of *n*-body integrations, using the *Katana* supercomputing cluster at the University of New South Wales, and the *Epic* supercomputing cluster, now retired, hosted by iVEC in Western Australia. The three suites of integrations are setup, where possible, in identical manners. They ran for a period of ten million years in to the future, from initial conditions based on the NASA DE431 ephemeris [62]. In order to maximise the accuracy of the simulations, an integration time-step of 1 day was used. The orbital elements of each of the planets are output at 1,000-year intervals for every simulation.

The first, and most detailed, suite of integrations was that which examined the influence of Jupiter's orbit on Earth's Milankovitch cycles. For these simulations, the preliminary results of which we presented in [54], we tested 159,201 unique versions of our Solar System. In those simulations, the initial orbits of all the planets except Jupiter were held fixed at their DE431 values. By contrast, the semi-major axis and eccentricity of Jupiter's orbit was varied systematically through the runs, such that a 399 x 399 grid of solutions were tested in semi-major axis – eccentricity space. The semi-major axis values tested for Jupiter spanned a region of 4 au, centred on the DE431 value of $a = 5.203102$ au, and were distributed evenly throughout the range. The eccentricities considered were evenly distributed between circular orbits ($e = 0.0$) and those with moderate eccentricity ($e = 0.4$).

Given the wide range of Jovian orbits considered, we anticipated that at least some of the solutions tested would prove to be dynamically unfeasible – as has often be found when testing the proposed orbits of newly discovered exoplanetary systems (e.g. [63][64][65]). To address this issue, if any of the planets collided with one another, fell into the Sun, or reached a barycentric distance of 40 au, the simulations were halted, and the time at which the event happened was recorded.

These simulations allowed us to create maps showing how the Earth's orbital elements varied with time as a function of Jupiter's initial orbital semi-major axis and eccentricity, as well as allowing us to map out the regions of instability where Jupiter tore the Solar System asunder. These maps build on our earlier work studying the stability of Solar System and exoplanetary orbits (e.g. [66][67]), and provide a visual guide to the manner in which Earth's Milankovitch cycles are modified by Jupiter's orbital characteristics.

We have now complemented these simulations of Jupiter's influence by examining the influence of Mars and Venus on the Earth's orbital evolution. These new simulations each considered 39,601 different orbits for the planet in question. For both planets, we tested 199 unique values of eccentricity spread evenly between 0.0 and 0.3. At each of these eccentricities, we tested 199 unique values of semi-major axis, centred on the DE431 values for the planet's orbit. For Mars, the semi-major axes chosen were therefore centred on 1.524 au, and ranged as far as 0.45 au on either side of that value (i.e. ranging from 1.074 To 1.974 au). For Venus, the semi-major axes centred on 0.723 au, and ranged 0.277 au on either side of that value (for a range between 0.446 and 1.000 au).

We are currently in the process of taking the numerical results of these simulations (the orbital elements of the Earth across the three suites of integrations) and using them as input for simple climate models (e.g. [68]). Those calculations for our Jupiter simulations are almost complete, and we anticipate that the full analysis will be ready in the near future.

## Preliminary results

In order to directly compare the influence of Jupiter, Mars and Venus on Earth's Milankovitch cycles, we present here the root-mean-square rates of change of the Earth's orbital inclination and eccentricity over the ten million year period integrated in this work. Regions plotted in black are those where the Solar System was rendered unstable within the ten million year timeframe of our simulations by the initial architecture chosen. Figure 1 presents the rate of change of Earth's orbital eccentricity as a function of the chosen planet's semi-major axis and eccentricity, whilst Figure 2 presents the rate of change of its inclination with the planet's semi-major axis and eccentricity.

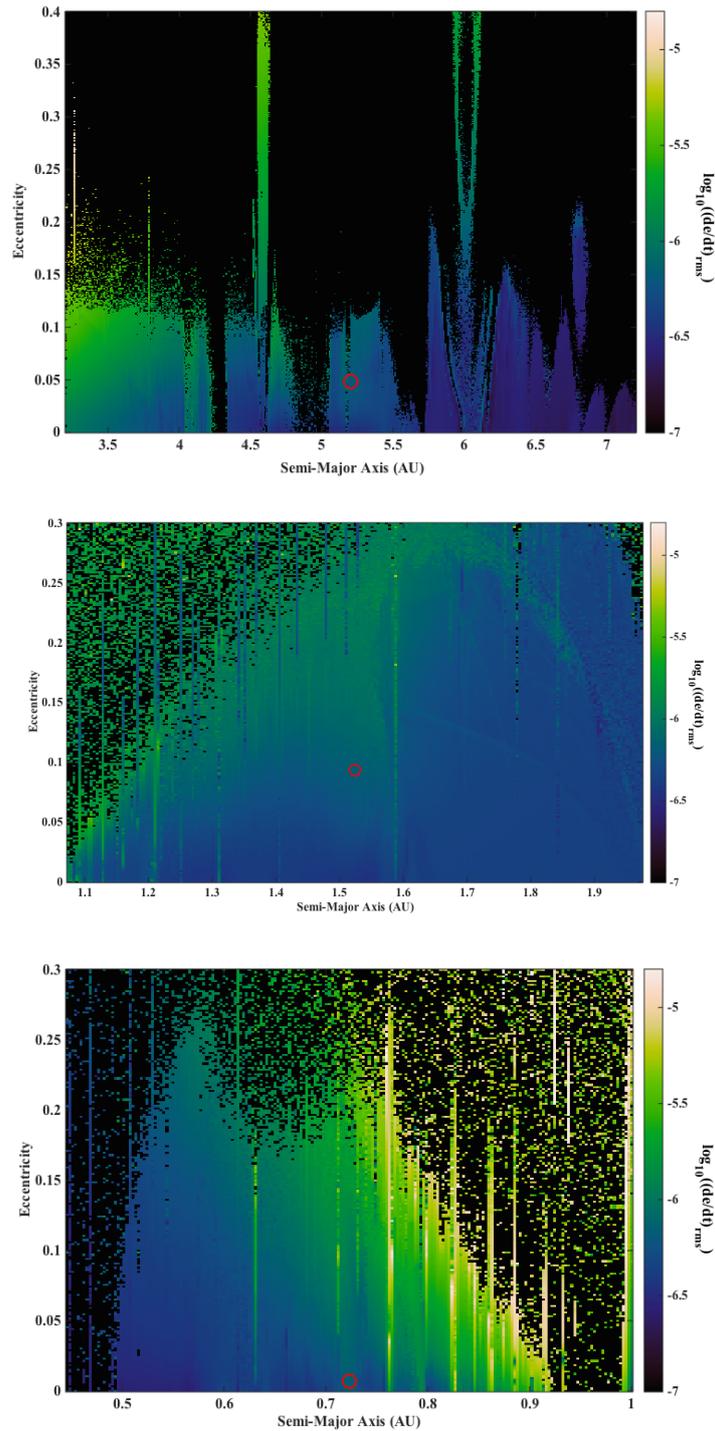

*Figure 1: The root-mean-square rate of change of Earth's orbital eccentricity over a ten million year period, plotted on a logarithmic scale, to illustrate the influence of the Solar System's architecture on our planet's Milankovitch cycles. The top panel shows the variability as a function of Jupiter's initial semi-major axis and eccentricity, with the variability as a function of Mars and Venus' orbits being plotted in the centre and lower panels, respectively. Regions in black are those for which the Solar System was rendered dynamically unstable by the chosen architecture, with planets either colliding with one another, being thrown into the Sun, or reaching a barycentric distance of 40 au (for Jupiter) or 1000 au (Venus and Mars). The orbit of the planet in our own Solar System lies at the centre of the hollow red circle.*

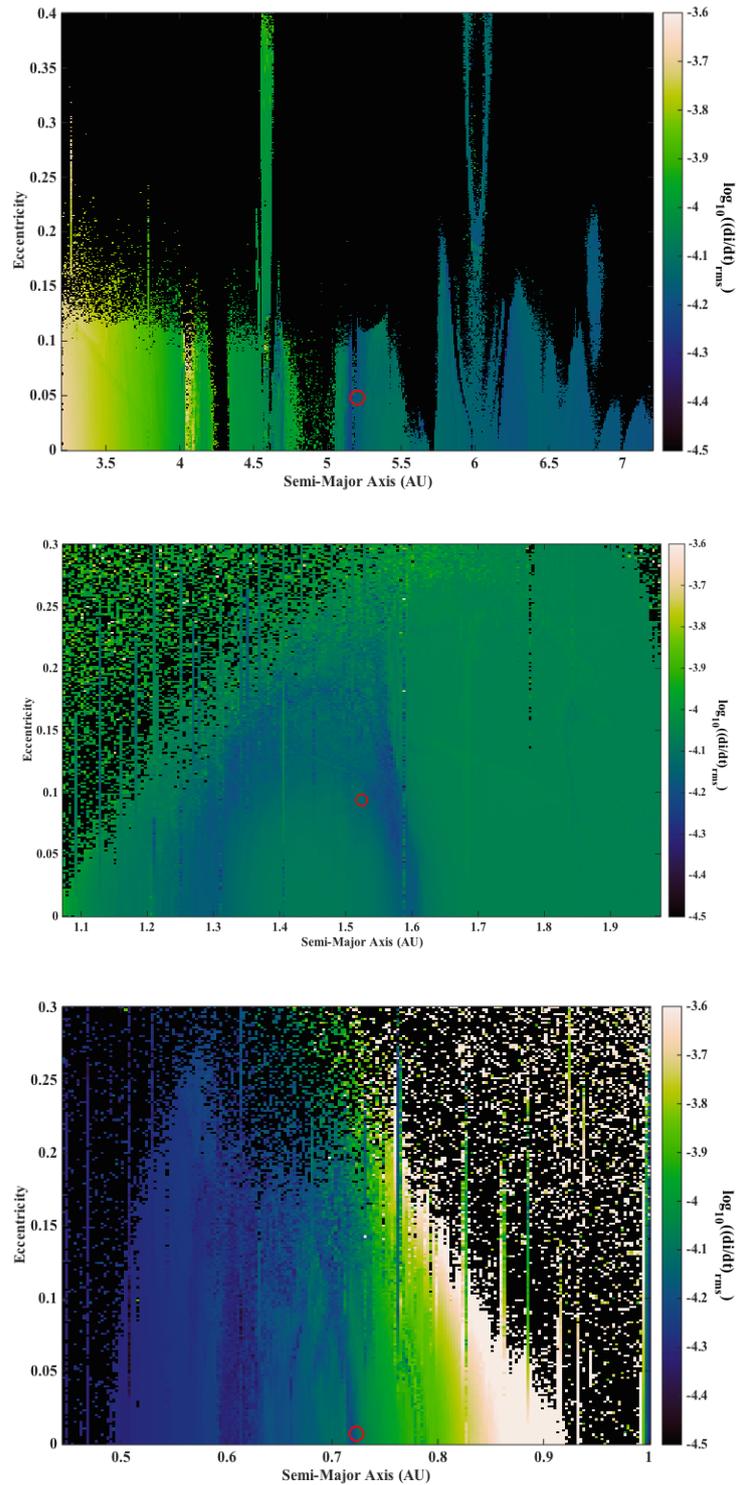

*Figure 2: The root-mean-square rate of change of Earth's orbital inclination over ten million years, on a logarithmic scale, as a function of the initial orbit of Jupiter (top), Mars (middle) and Venus (bottom). The regions shown in black are those where the Solar System become unstable within the ten million year runtime, with planets being ejected, colliding with one another, or being thrown into the Sun. As before, the hollow red circle denotes the orbit of the planet in question in our own Solar System.*

## Discussion and Conclusions

Several things are immediately apparent when one compares the influence of Jupiter, Mars and Venus on the Earth's orbital evolution (Figures 1 and 2). The first is how moving the three planets impacts the stability of the Solar System. As we noted in [54], the Solar System's stability is strongly dependent on Jupiter's initial orbit, with only a relatively small change in the planet's orbit being enough to destabilise the entire Solar System.

The same, to an extent, is true of Venus – if placed on an orbit with a perihelion distance less than ~0.485 au, Venus comes close enough to Mercury that the two undergo strong gravitational interaction, destabilising the Solar System. Similarly, if Venus' aphelion exceeds ~0.92 au, it strongly interacts with the Earth, again leading to the Solar System's disintegration. This result is not hugely surprising. Indeed, [69] studied the effect of tightly-packing planets on their orbital stability, and found that two planet systems are typically stable so long as the planets therein are separated by more than 2√3 times their mutual Hill radius, where for low eccentricity orbits the mutual Hill radius is defined as:

$$R_H = \left[\frac{(m_1 + m_2)}{3M}\right]^{1/3} \left[\frac{(a_1 + a_2)}{2}\right] \quad [1].$$

Here $m_1$ and $m_2$ are the masses of the two planets in question, $M$ is the mass of the central star, and $a_1$ and $a_2$ are the semi-major axes of the orbits of the two planets. In the case of Earth and Venus, we find that when Venus is at 0.92 au from the Sun, with the Earth at perihelion at ~0.98 au, their *instantaneous* mutual Hill radius is 0.012 au. Therefore, the two planets are separated by just approximately five mutual Hill radii, only slightly more than the 2√3 Hill radii noted by [69]. Accordingly, the boundary between unstable and stable regimes in the region ~0.75 – 0.92 au tends to follow the line of constant aphelion distance meeting this criterion. It should be noted, too, that the values plotted for the semi-major axis and inclination of Venus (and Mars and Jupiter, in the other plots), are the instantaneous values at the start of the simulations – and it is quite feasible that planets which start out separated by more than these 2√3 Hill radii will occasionally experience periods that offer encounters closer than this distance. For this reason, the boundary between stable and unstable regimes does not perfectly follow a line of constant Venutian aphelion, but instead tends along that line, with a certain amount of noise and fluctuation.

An exception to these regions of stability and instability resulting from close approaches occurs when Venus is placed at the location of mean-motion resonances with either Earth or Mercury, which can be seen as the striking vertical features in both figures (albeit most pronounced in Figure 1). Perhaps the most interesting of these regions of resonant stability is the one in the vicinity of the Earth's orbit, around 1 au from the Sun. Here, Venus and Earth can remain dynamically stable, and are most likely trapped in mutual 1:1 resonance – in other words, Venus is acting as a Trojan companion to the Earth. Such companions are widely observed in our own Solar System (e.g. the Jovian and Neptunian Trojans, [70][71][72][73]), and the existence of planet-mass Trojans has been suggested as a means by which potentially habitable worlds could be found in planetary systems with hot- or warm-Jupiters (e.g. [74][75][76]).

Such papers usually discuss the existence of Trojan companions that are far less massive than the planet with which they share an orbit – just as is the case for the Trojan companions to the

Solar System's planets. However, [77] considered the question of whether planets of similar or equal mass could survive as mutual Trojan companions. They found that mutually resonant planets could remain dynamically stable so long as their combined mass did not exceed ~1/26$^{th}$ of the mass of their host star. As such, our result that Venus and Earth *could* form a dynamically stable Trojan couple is not perhaps that surprising – but it is certainly interesting to consider, and opens up an additional direction for further study in the future. Namely – would the existence of a Trojan companion render the climate of a given planet too unstable for the evolution of detectable life? Whilst our results offer no definitive conclusions to the answer to this problem, it is clear that the stable 1:1 resonant solutions in our simulations featured variability in orbital inclination and eccentricity far more rapid than are experienced within our current Solar System. For this case, at least, it might therefore be to consider those solutions less habitable than our own Earth – but significantly more work would be required to extend this to a general argument for all Trojan scenarios.

In the case of Mars, far less instability is seen. Once again, though, it is clear that moving Mars too close to the orbit of the Earth will render the system unstable (region leftward of a line of constant perihelion ~1.06 au). Even above this line, though, a significant number of solutions survive for the 10 Myr of our simulations. This might well be an artefact of the 1000 au ejection criteria used in the simulations for Mars, however – and we intend to re-run those simulations in the near future using a 40 au ejection distance to match our work with Jupiter.

The influence of mean-motion resonances can be clearly seen in the plots detailing Mars' influence. To the outer edge of those plots, unstable solutions are found near a = 2 au. This instability is the result of the influence of the 4:1 mean-motion resonance with Jupiter, which is centred at 2.06 au, and is the cause of a broad Kirkwood gap in the asteroid belt (e.g. [78]). The influence of a less disruptive mean-motion resonance can be seen at ~1.587 au, in the form of a vertical stream of slightly accelerated variability in Earth's eccentricity evolution (Figure 1, middle plot). This is the location of the 2:1 mean-motion resonance between Earth and Mars.

Whilst the effect of resonances between Earth and Mars is somewhat understated in both Figure 1 and Figure 2, resonant interactions between Earth and Venus have a far stronger influence on the rate of change of Earth's orbit. This is not unsurprising – Venus is almost eight times more massive than Mars, and not hugely less massive than the Earth – so with all else equal, one would expect it to more strongly perturb our planet. This is borne out in the strength of the resonant perturbations on Earth visible in the lower panel of Figure 1, and the broader region of instability caused by Venus and Earth undergoing close approaches. Mars, the less massive of the two planets, can approach Earth slightly closer than Venus before destabilising the Solar System.

Beyond these observations on broad stability, a further surprising result is clear from our simulations. When moved away from its current location, Venus can induce far more rapid shifts in both Earth's orbital eccentricity (Figure 1) and inclination (Figure 2) than are observed for our simulations of Jupiter's influence. Before we leap to the conclusion that Venus actually has a greater influence on Earth's climate than does Jupiter, we need here to include the caveat that, even in the most extreme scenarios tested, Jupiter remained far beyond Earth's orbit, at ~3.2 au from the Sun. It would be interesting to see how Jupiter would affect Earth's inclination and eccentricity were it moved further inward – though obviously it would eventually destabilise the system were it moved sufficiently far inward!

Whilst the influence of Venus on Earth's eccentricity evolution is strongest when resonant interactions come into play, it is interesting to observe that the same is not true for its influence on Earth's orbital inclination. The more closely Venus and Earth approach one another, the more rapidly the inclination of Earth's orbit is driven to change. This behaviour is strikingly illustrated towards the right-hand edge of the stability region in the lower panel of Figure 2.

Aside from those regions where close approaches or resonant interactions begin to render the Solar System unstable, our results show that Mars has a far less significant impact on the orbital evolution of the Earth than either Venus or Jupiter. The plots in Figures 1 and 2 all span the same range in inclination and eccentricity variability – and the variation seen across the bulk of the Martian simulations is lower than that for either Venus or Jupiter. Mean motion resonances (vertical strips in the plots) do induce some variability, and the eccentricity and inclination plots also reveal subtle variations – the 'breaking wave' in Figure 1, and the prominence-like loop in Figure 2. These are most likely the result of secular interactions between the planets, but are relatively minor features compared to the strong variability resulting from the other two planets.

In the coming year, we intend to pair our simulations of Venus, Mars and Jupiter's influence with climate models, which will incorporate the precession of the Earth's axis, allowing us to determine the variability of the Earth's obliquity in our simulations. The obliquity variability is the key dynamical parameter in translating the ongoing perturbations on Earth's motion into the resulting climatic variability, and will feed directly into our climate models.

Beyond that work, we have begun a suite of integrations that will examine how Earth's initial orbit influences its Milankovitch cycles. In those simulations, we hold the initial orbits of the other planet fixed, as described in this work, and instead move the Earth within the Solar System. Those simulations are currently underway on UNSW Australia's *Katana* supercomputing cluster, and will bring the first phase of our study of planetary architecture and climate stability to a close.

Taken in concert, our work will help us to uncover the degree to which fine tuning in planetary architecture can impact the climate, and potential habitability, of planets like the Earth. In doing so, it will give us a feel for how 'unusual' the Earth's climatic variability is. A core tenant of the 'Rare Earth' philosophy (e.g. [79]) is that the Earth is unusually fortunate in its properties – almost a fluke of nature that just happens to be a safe haven in an inimical cosmos. It suggests that, for example, the impact rate at Earth is far lower than it would be were the giant planet Jupiter not present – in other words, that without our good fortune in having a giant protector, we would not be here. Whilst that particular aspect of the hypothesis has since been effectively demolished (see e.g. [56][80][81][82][83][84]), the idea underpinning the thesis remains under debate. Is the Earth unusually quiescent? Would most other 'Earths' be far less habitable than our own? Once we bring together climate modelling with our *n*-body simulations, we will begin to get a feel for just how precarious is Earth's climate, in the context of the architecture of our own planetary system.

Beyond this, however, our work will build a framework by which the climatic variability of newly discovered exoEarths can be estimated – once such planets are found. Whilst this is far from the only consideration when it comes to determining the suitability of those planets as targets in the search for life, it will still provide a useful datum – potentially helping to rule out some otherwise promising targets, allowing observers to instead focus their efforts at more promising worlds.

## Acknowledgements

JH is supported by USQ's Strategic Research Fund: the STARWINDS project. The work was supported by iVEC through the use of advanced computing resources located at the Murdoch University, in Western Australia. The authors wish to thank A/Prof Ramon Brasser and the anonymous second referee of this work, whose feedback greatly helped to improve the clarity and flow of the paper.